\documentclass[aps,prb,twocolumn,floatfix,showpacs]{revtex4}
\usepackage{amsmath,amssymb,graphicx,bm}
 \def\veck{\mathbf k}
\def\vecq{\mathbf q}

\begin{document}
\title{Thermodynamically consistent description of criticality in models of correlated electrons}

\author{V\'aclav  Jani\v{s} }\author{Anna Kauch} \author{Vladislav Pokorn\'y} 

\affiliation{Institute of Physics, The Czech Academy of Sciences, Na Slovance 2, CZ-18221 Praha  8,  Czech Republic}
\email{janis@fzu.cz}


\date{\today}


\begin{abstract}
Criticality in models of correlated electrons emerges in proximity of a low-temperature singularity in a two-particle Green function. Such singularities  are generally related to a symmetry breaking of the one-particle self-energy. A consistent description demands that the symmetry breaking in the self-energy emerges at the critical point of the respective two-particle function. This cannot easily be achieved in models of correlated electrons, since there are two ways connecting one- and two-electron functions that cannot be made fully equivalent in approximations. We present a general construction of  diagrammatic two-particle approximations consistent with the one-particle functions so that both produce  qualitatively the same quantum critical behavior in thermodynamically equivalent descriptions. The general scheme is applied on the single-impurity Anderson model to derive qualitatively the same Kondo critical scale from the spectral function  and the magnetic susceptibility.      
\end{abstract}

\maketitle 

\section{Introduction}
\label{sec:Intro}

A consistent and reliable description of the low-temperature behavior of strongly correlated electron systems has not yet been reached in spite of decades of intensive research in this field. Most properties of weakly and moderately coupled electrons in metals are captured in a sufficient extent by the Fermi-liquid theory. The problems arise when one tries to extend Fermi-liquid solutions to the strong-coupling regime. Fermi liquid becomes unstable due to quantum dynamical fluctuations and the electron system approaches a quantum critical point for a sufficiently strong interaction.  Non-perturbative approaches are then needed. One option is to use unbiased numerical simulations that are not restricted to weak coupling. Various variants of quantum Monte Carlo,\cite{Landau:2014aa,Gull:2011aa} or numerical,\cite{Bulla:1998aa,Bulla:2008aa}  density matrix\cite{Schollwock:2005aa,Wolf:2015aa} or functional\cite{Kopietz:2010ab,Karrasch:2008aa} renormalization group, are widely used to obtain quantitative non-perturbative results in the whole range of the interaction strength. Monte Carlo simulations use Matsubara formalism and are good for relatively high temperatures. Numerical works based on renormalization group schemes do well for low-lying energy states at low temperature.  In both cases only static thermodynamic properties are directly available. Numerical solutions are restricted to finite-size clusters and relatively small sets of Matsubara frequencies or low-lying energy states. Moreover, they address primarily one-electron functions. Proximity of the critical points demands, however, controlling two-particle and response functions that, unlike the one-particle ones,  may diverge at the critical point. The singularities in two-particle functions and the critical, non-analytic behavior can be identified and fully controlled only analytically. The singularities and divergencies must be treated separately from the non-divergent quantities so that to reach  stable solutions  and  to avoid spurious behavior in numerical calculations and iterations.      

Analytic approaches are generally based on a many-body perturbation, diagrammatic expansion in the interaction strength.  They work well in the weak-coupling, Fermi-liquid regime. To extend them beyond weak electron correlations one needs to sum infinite series of specific classes of diagrams and make the approximations non-perturbative and self-consistent. Self-consistency cannot be introduced in an arbitrary manner, since there is a danger that some of the macroscopic conservation laws be broken. The canonical way how to achieve conserving and thermodynamically consistent approximations was outlined by Baym and Kadanoff.\cite{Baym:1961aa,Baym:1962aa} When an approximation can be derived from a generating Luttinger-Ward functional of the full one-particle propagator, the continuity equation holds and mass is conserved. The fundamental quantities in the Baym-Kadanoff construction are the generating functional, the one-particle propagator, and the self-energy. The two-particle vertex functions are not explicitly addressed and are determined via functional derivatives from the self-energy, treated as a functional of the renormalized one-electron propagator.  The renormalization in the Baym-Kadanoff construction does renormalize the bare interaction.\cite{Janis:1999aa} Consequently, there is no direct control of singularities in the Bethe-Salpeter equations for the two-particle functions. To circumvent this problem a so-called parquet scheme was introduced taking into account also renormalizations of two-particle vertices.\cite{Sudakov:1956aa,DeDominicis:1962aa,DeDominicis:1963aa,DeDominicis:1964aa,DeDominicis:1964ab} The renormalization of the unperturbed propagator and the bare interaction must be made in a coordinated way so that not to break macroscopic continuity equation.\cite{Bickers:1989aa}  Even if we guarantee mass conservation in the theory with renormalized one and two-particle functions, we are unable to match the irreducible vertex derived from the self-energy via the functional Ward identity with the full vertex used in the Schwinger-Dyson equation.\cite{Levine:1968aa,Janis:1998aa}   This inconsistency was first met already earlier when attempting to go beyond the weak-coupling theory of superconductivity of Bardeen, Cooper and Schrieffer (BCS).\cite{Kadanoff:1961aa} When the fully self-consistent T-matrix approximation (TMA), conserving in the Baym-Kadanoff sense, is used to renormalize multiple electron-electron scatterings of the BCS theory, the pole in the two-particle correlation function does not lead to opening of the superconducting gap.\cite{Bishop:1974aa}  To resolve the problem a variety of modifications of TMA combining self-consistent and bare propagators have been proposed.\cite{Chen:2005aa,Lipavsky:2008aa} These attempts to resolve the inconsistency of the fully self-consistent TMA  are mostly ad hoc suggestions that lack the solid basis on which one could systematically build up further  improvements. A more general, internally consistent scheme leading to a reliable description of quantum criticality is needed.     

The aim of this paper is to address the problem of a systematic description of  criticality in correlated electron systems. Critical behavior there is indicated by a singularity in a Bethe-Salpeter equation. The principal new idea of our construction is to use the irreducible vertex from the singular Bethe-Salpeter equation, instead of the self-energy, as a generating function obtained from a diagrammatic perturbation  expansion. We solve a linearized Ward identity for the given irreducible vertex to determine a thermodynamic self-energy.  We  introduce self-consistency into the diagrammatic expansion by renormalizing the one-particle propagators by the thermodynamic self-energy. In this way we achieve a consistency between the divergence in the Bethe-Salpeter equation and the corresponding symmetry breaking in the thermodynamic self-energy. Approximate irreducible  vertices do not, however, guarantee that this thermodynamic self-energy fulfills the Schwinger-Dyson equation. It can actually be achieved only in the exact dynamical solution for the irreducible vertex that is far beyond reach. The physical self-energy is nevertheless the one fulfilling the Schwinger-Dyson equation. We hence introduce another self-energy from the Schwinger-Dyson equation that we call spectral.  The one-particle propagators in the Schwinger-Dyson equation do not change their renormalization in this definition. It means that the spectral self-energy no longer enters a self-consistency loop. The spectral self-energy is the principal output of the theory and is used to determine all physical and measurable quantities. In this way the corresponding symmetry of the one-electron propagator with the spectral self-energy gets broken at the critical point of the Bethe-Salpeter equation and the respective response function.

The presentation of our construction consists of four hierarchical levels.  In the first one we identify the origin of the failure of the $\Phi$-derivable approximations  to match the symmetry breaking in the self-energy with the singularity in a Bethe-Salpeter equation  (Sec.~\ref{sec:thermodynamic}). In the following step, Sec.~\ref{sec:2P-consistent}, we introduce the construction of the thermodynamic self-energy from the electron-hole irreducible vertex via the Ward identity and its utilization in the construction of physical quantities near a magnetic phase transition.  One needs to introduce approximations on the two-particle vertex to reach quantitative results. We use the parquet approach to achieve a two-particle self-consistency in the perturbation expansion for two-particle vertices in Sec.~\ref{sec:2P-SC}. Finally, we choose the single-impurity Anderson model (SIAM) to demonstrate explicitly how our construction with the simplest  self-consistent approximation for the irreducible vertex leads to qualitatively the same Kondo scale  in the spectral function and in the magnetic susceptibility, Sec.~\ref{sec:SIAM}. Numerical calculations for SIAM at half-filling are presented in Sec.~\ref{sec:numeric}. In Sec.~\ref{sec:discussion} we discuss consequences and changes our construction brings compared to the Baym and Kadanoff approach.

\section{Conserving approximations: Generating functional and charge conservation}
\label{sec:thermodynamic}

We use the paradigm Hubbard Hamiltonian of interacting lattice electrons 
\begin{align}
\widehat{H}_{\mu}&=\sum_{{\bf k}\sigma} \epsilon({\bf k})
   c^{\dagger}_{{\bf k}\sigma}
  c^{\phantom{\dagger}}_{{\bf k}\sigma}   +
  U\sum_{{\bf i}}\widehat{n}_{{\bf i}\uparrow}\widehat{n}_{{\bf i}
    \downarrow} -\sum_{\mathbf{i}\sigma}\mu_{\sigma}\widehat{n}_{{\bf i}\sigma} 
\end{align}
allowing us to study strong electron correlations non-perturbatively.  We denoted $\epsilon(\veck)$ the lattice dispersion relation, $\mu_{\sigma}= \mu + \sigma h$ is the spin-dependent chemical potential with a  magnetic field $h$.  Operators $ c^{\dagger}_{{\bf k}\sigma}$, $c^{\phantom{\dagger}}_{{\bf k}\sigma}$ create and destroy electron with quasi-momentum  $\veck$ and $\widehat{n}_{{\bf i}\sigma}$ is the operator of the electron density on site $\mathbf{R}_{i}$. The fundamental ingredients of the description of models of interacting fermions are  Green functions for which we introduce a perturbation (diagrammatic) expansion. We use a renormalized perturbation expansion in the interaction strength and represent it with the aid of Feynman diagrams.  

It is generally believed that the so-called $\Phi$-derivable approximations with the generating Luttinger-Ward functional $\Phi[G]$ are thermodynamically consistent and obey all conservation laws. This construction leads to the one-electron irreducible  vertex (self-energy) as the principal object of the perturbation expansion. If we want to control directly also the two-particle irreducible vertices it is necessary to distinguish electron and hole propagators that we denote $G$ and $\overline{G}$, since they generate different types of the two-particle irreducibility. The generating functional for independent electron and hole functions then reads  
\begin{multline}\label{eq:PhiG}
 \frac 2{{ N}}\Omega [\Sigma ,G,\overline{\Sigma},\overline{G}]  =\Phi [U;G,\overline{G}]
 \\
   - \frac 1{\beta {N}}\sum_{\sigma n,{\bf k}} \left\{      
 e^{i\omega_n0^{+}}\ln \left[ i\omega _n+\mu _\sigma -\epsilon  
     ({\bf k}) -\Sigma_\sigma ({\bf
       k},i\omega_n)\right] 
          \right. \\  \left. 
           +  e^{-i\omega_n0^{+}}\ln \left[- i\omega _n+ \mu _\sigma - \epsilon  
     (-{\bf k}) -\overline{\Sigma}_\sigma (-{\bf
       k},-i\omega_n)\right]
       \right. \\  \left. 
       +\ G_\sigma ({\bf k},i\omega_n)\overline{\Sigma}_\sigma (-{\bf k},-i\omega_n) 
        \right. \\  \left. 
       + \overline{G}_\sigma (-{\bf k}, -i\omega_n)\Sigma _\sigma ({\bf k},i\omega_n) 		\right\}  \,,
\end{multline}
where $N$ is the number of lattice sites, $\beta = 1/k_{B}T$, and $\overline{\Sigma}$  is the hole  self-energy.  We set $k_{B}=1$. In equilibrium the electron-hole symmetry imposes the following relations  $\overline{\Sigma}_{\sigma}({\bf k},i\omega_n) = \Sigma_{\sigma}(-{\bf k},-i\omega_n)$, and $\overline{G}_{\sigma}({\bf k},i\omega_n) = G_{\sigma}(-{\bf k},-i\omega_n)$ that we use in final expressions. We distinguish the electron and hole functions only to derive equations for the one and two-particle irreducible vertices being (functional) derivatives of the Ward-Luttinger  functional $\Phi[U;G,\overline{G}]$. 

\begin{figure} 
\hspace*{-10pt}\includegraphics[width=9cm]{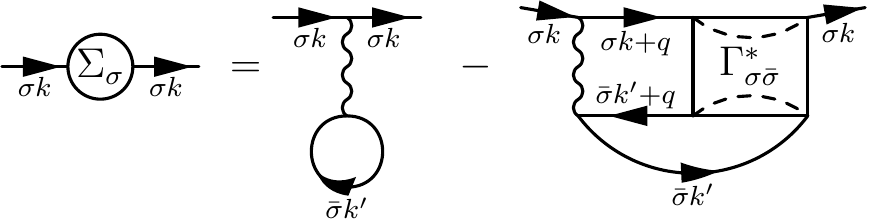} 
\caption{Diagrammatic representation of the Schwinger-Dyson equation,~\eqref{eq:SDE-symbolic}. The hole Green function $\overline{G}_{\bar{\sigma}}$ propagates charge  from right to left, as indicated by arrow. We denoted $\bar{\sigma} = - \sigma$ and $k=(\veck,i\omega_{n})$, $q=(\vecq,i\nu_{m})$.   \label{fig:SDE}}
\end{figure}

The first fundamental equation is that for the self-energy of the Hubbard model 
\begin{multline}\label{eq:SDE-symbolic}
\Sigma_{\sigma}[U;G,\overline{G}] = \frac{\delta \Phi[U;G,\overline{G}]}{\delta \overline{G}_{\sigma}} 
=  U\left\langle \overline{G}_{-\sigma}\right\rangle
\\
 - U G_{\sigma}\overline{G}_{-\sigma}\star\Gamma^{\ast}_{\sigma-\sigma}[U;G,\overline{G}] \circ G_{-\sigma} 
\end{multline}
and has the form of the Schwinger-Dyson equation, see Fig.~\ref{fig:SDE}.  The angular brackets stand for the sum over momenta and Matsubara frequencies.  Notice that the linear, Hartree term on the right-hand side can be represented equivalently also with the particle propagator $U\langle G_{-\sigma}\rangle$. The static local function cannot distinguish between the particle and the hole.  We skipped the momentum and frequency variables but distinguished  electron-hole (antiparallel lines),  $\star$, and electron-electron (parallel lines), $\circ$, propagation corresponding to different way of attaching the frequencies and momenta. We denoted $\Gamma^{*}$ a two-particle vertex. The one-particle propagators are determined from the Dyson equation, $G_{\sigma}({\bf k},i\omega_n) = 1/\left[ i\omega _n+\mu _\sigma -\epsilon({\bf k}) -\Sigma_\sigma ({\bf k},i\omega_n)\right]$ resulting from a stationarity condition $\delta\Omega[\Sigma ,G,\overline{\Sigma},\overline{G}]/\delta\overline{\Sigma}(-\veck,-i\omega_{n}) = 0$ and the electron-hole symmetry.
   
The full vertex is generally represented via irreducible vertices and Bethe-Salpeter equations. The two-particle irreducibility is not uniquely defined, hence there are several independent Bethe-Salpeter equations. As an example we choose the singlet electron-hole Bethe-Salpeter equation for the two-particle vertex $\Gamma_{\sigma-\sigma}$
\begin{multline}\label{eq:BSE-eh-symbolic}
\Gamma_{\sigma-\sigma}[U;G,\overline{G}] = \Lambda^{eh}_{\sigma-\sigma}[U;G,\overline{G}]
\\
  - \Lambda^{eh}_{\sigma-\sigma}[U;G,\overline{G}]G_{\sigma}\overline{G}_{-\sigma}\star\Gamma_{\sigma-\sigma}[U;G,\overline{G}]  
\end{multline}
where the irreducible vertex $\Lambda^{eh}_{\sigma-\sigma}[U;G,\overline{G}]$ should be connected in the conserving theory with  the self-energy via a functional Ward identity\cite{Baym:1962aa}
\begin{equation}\label{eq:WI-functional}
\Lambda_{\sigma-\sigma}^{eh}[U;G,\overline{G}] =  \frac{\delta \Sigma_{\sigma}[U;G,\overline{G}]}{\delta \overline{G}_{-\sigma}} \,. 
\end{equation}
This relation can easily be proved from the diagrammatic representation of the Schwinger-Dyson equation in Fig.~\ref{fig:SDE} and the diagrammatic definition of the electron-hole irreducibility.

Analogously, we can represent the full vertex $\Gamma_{\sigma-\sigma}$ via the irreducible vertex in the electron-electron scattering channel  $\Lambda^{ee}_{\sigma-\sigma}= \delta \Sigma_{\sigma}/\delta G_{-\sigma}$.  There are also triplet vertices  $\Lambda^{ee}_{\sigma\sigma}= \delta \Sigma_{\sigma}/\delta G_{\sigma}$ and $\Lambda^{eh}_{\sigma\sigma}= \delta \Sigma_{\sigma}/\delta \overline{G}_{\sigma}$  determining the full triplet vertex $\Gamma_{\sigma\sigma}$ that we do not consider here. Notice that to distinguish different two-particle irreducibilities we have to distinguish electrons from holes in the functional derivatives. 

The two vertex functions from the Schwinger-Dyson and Bethe-Salpeter equations, Eqs.~\eqref{eq:SDE-symbolic} and~\eqref{eq:BSE-eh-symbolic} equal in the exact theory. That is $\Gamma^{\ast}[U;G,\overline{G}]  = \Gamma[U;G,\overline{G}]$. This cannot, however, be achieved in accessible approximate treatments.  Consistency between the one and two-particle functions in the Baym-Kadanoff construction with the generating self-energy functional  $\Sigma[G]$  is guaranteed if the functional Ward identity, Eq.~\eqref{eq:WI-functional}, is obeyed and the full vertex is represented via the Bethe-Salpeter equation~\eqref{eq:BSE-eh-symbolic}. If $\Gamma^{\ast}[U;G,\overline{G}]  = \Gamma[U;G,\overline{G}]$ then a functional derivative of the self-energy and of the irreducible vertex must comply with another equation  when Eq.~\eqref{eq:BSE-eh-symbolic} is inserted in Eq.~\eqref{eq:SDE-symbolic}
\begin{multline}\label{eq:SDE-derivative}
\frac{\delta \Sigma_{\sigma}[U;G,\overline{G}]}{\delta \overline{G}_{-\sigma}} =  U - U\left[ 1 + G_{\sigma}\overline{G}_{-\sigma}\Lambda^{eh}_{\sigma-\sigma}\star\right]^{-1}G_{\sigma}\left\{\Lambda^{eh}_{\sigma-\sigma}\phantom{\frac 12}
\right. \\ \left.
 +\ \overline{G}_{-\sigma}\frac{\delta\Lambda^{eh}_{\sigma-\sigma}}{\delta \overline{G}_{-\sigma}}\right\}
\left[ 1 + \star G_{\sigma}\overline{G}_{-\sigma}\Lambda^{eh}_{\sigma-\sigma}\right]^{-1}\circ G_{-\sigma} \,.
\end{multline}
It is evident that Eqs.~\eqref{eq:WI-functional} and~\eqref{eq:SDE-derivative} cannot be obeyed simultaneously with approximate irreducible functions and $\Gamma^{\ast}[U;G,\overline{G}]  \neq \Gamma[U;G,\overline{G}]$.  To keep the approximation for the self-energy from Eq.~\eqref{eq:SDE-symbolic} conserving one has to treat vertex $\Gamma^{\ast}$  as an auxiliary function and to give the physical meaning only to vertex $\Gamma$ from the Bethe-Salpeter equation~\eqref{eq:BSE-eh-symbolic}. This is, however, possible only up to a critical point, divergence in the auxiliary vertex $\Gamma^{\ast}$. One cannot continue the approximation beyond this critical point unless the Ward identity is obeyed. At least to the extent that would guarantee that the critical point in the vertex  function $\Gamma^{\ast}$ introduces a symmetry breaking and emergence of an order parameter in the self-energy.   The problem is more severe, since the two vertex functions $\Gamma$ and $\Gamma^{\ast}$ in the $\Phi$-derivable approximate theories lead to two different critical points that should coincide in the exact solution.
 %

The existence of two vertex functions $\Gamma^{\ast}$ and $\Gamma$ is a consequence of the inability to obey charge conservation. This can be demonstrated on a singlet correlation function $\mathcal{C}_{\mathbf{ij}} =  \langle \widehat{n}_{{\bf i}\uparrow }\widehat{n}_{{\bf j}\downarrow}\rangle - \langle \widehat{n}_{{\bf i}\uparrow}\rangle  \langle \widehat{n}_{{\bf j}\downarrow}\rangle$. We can generate its dynamical version in two ways by functional derivatives with respect to space and time inhomogeneous perturbations $U\rightarrow U+\delta
U_{\bf ij}(\tau,\tau')$ and $\mu_\sigma\to\mu_\sigma+ \delta\mu_{{\bf
    i}\sigma}(\tau)$ as follows\cite{Janis:1998aa}
\begin{equation}
  \label{eq:WI-Charge}
  \frac{\delta\Phi[U,G]}{\delta U_{\bf ij}(\tau,0)}\bigg|_{\delta U=0\atop
    \delta\mu=0} - \langle \widehat{n}_{{\bf i}\uparrow}\rangle \langle \widehat{n}_{{\bf i}\downarrow}\rangle 
    = - \frac{\delta
    G_{{\bf ii}\uparrow}(\tau,\tau^+)}{\beta\delta\mu_{{\bf
        j}\downarrow}(0)}\bigg|_{\delta U=0\atop\delta\mu=0} 
\, .      
\end{equation}
The  correlation function on the  left-hand side is constructed from vertex $\Gamma^{\ast}$ while that on the right-hand side from vertex $\Gamma$.  This dynamical equality expresses conservation of charge in the sense that the electrostatic interaction $U$ is generated entirely by the present charge densities. If the two definitions differ then there are additive sources of the particle interaction. Differences between the two definitions from Eq.~\eqref{eq:WI-Charge} become dramatic in criticality. They are unacceptable if both definitions produce different critical points. The left-hand-side function diverges at the critical point of the Bethe-Salpeter equation while the right-hand-side function diverges at the point where an order parameter in the self-energy emerges. A thermodynamically consistent description of quantum criticality must produce an unambiguous singularity in the two-particle vertices. 

The inability to fulfill  simultaneously Eqs.~\eqref{eq:WI-functional} and~\eqref{eq:SDE-derivative} is the origin of the problems of the self-consistent approximations in the Baym-Kadanoff approach.  The emergence of the order parameter, resulting from the divergence on the right-hand side of Eq.~\eqref{eq:WI-Charge}, is not matched by the divergence of the two-particle vertex from its left-hand side at the critical point.  We propose to relate the self-energy in the Green function on the right-hand side of Eq.~\eqref{eq:WI-Charge}  to the irreducible vertex from the left-hand side via the Ward identity, Eq.~\eqref{eq:WI-functional}, to reconcile the discrepancy.

\section{Two-particle approach: Thermodynamic consistency in criticality}
\label{sec:2P-consistent}

The starting point of our construction is the irreducible vertex of the critical Bethe-Salpeter equation that serves as a generating functional of the theory. It is  an input determined from a self-consistent perturbation expansion. The self-energy is then derived from this vertex via the Ward identity. This thermodynamically constructed  self-energy is used in the renormalized one-particle propagators. In this way we achieve consistency between the criticality in the two-particle vertices and derivatives of the one-particle self-energy. The diagrammatic perturbation expansion is hence not applied  on the one-particle self-energy but rather on the two-particle irreducible vertices that generate singularities in the Bethe-Salpeter equations. For the sake of simplicity we consider here only Bethe-Salpeter equations in the singlet electron-hole and electron-electron channels that we later use to form nontrivial approximations.   
 
\begin{figure}
\hspace*{-10pt}\includegraphics[width=9cm]
   {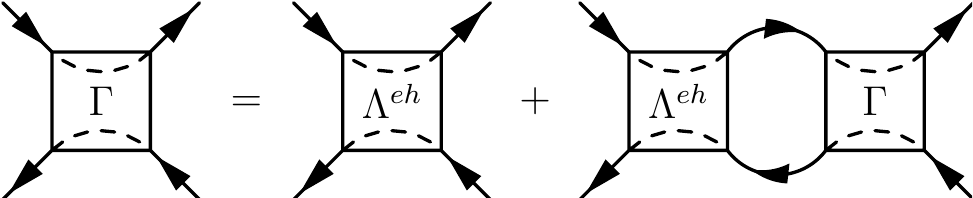}
   \caption{Diagrammatic representation of the Bethe-Salpeter equation,~\eqref{eq:BS-eh}, in the electron-hole channel summing electron-hole ladders. The dashed lines within the boxes indicate how incoming and outgoing electrons are interconnected. \label{fig:BSE-eh}}
\end{figure}

\begin{figure}
\hspace*{-10pt}\includegraphics[width=9cm]
   {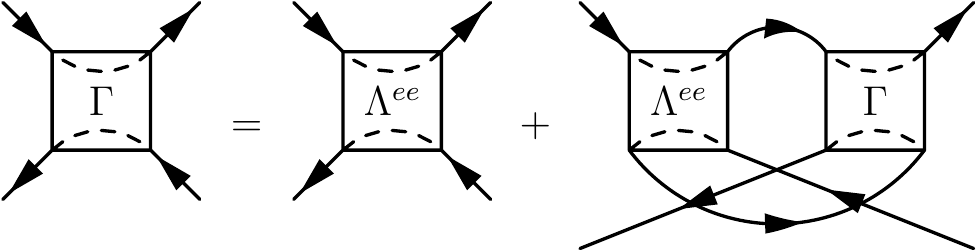}
   \caption{Diagrammatic representation of the Bethe-Salpeter equation,~\eqref{eq:BS-ee}, in the electron-electron channel summing electron-electron ladders. \label{fig:BSE-ee}}
\end{figure}

The Bethe-Salpeter equation for the multiple scatterings of the electron with spin up and the hole  with spin down, electron-hole singlet channel, reads 
\begin{multline}\label{eq:BS-eh}
\Gamma_{\uparrow\downarrow}(k,k';q) = \Lambda_{\uparrow\downarrow}^{eh}(k,k';q)
\\
  - \frac 1{\beta N}\sum_{k''} \Lambda_{\uparrow\downarrow}^{eh}(k,k'';q) G_{\uparrow}(k'') 
  \\
  \times G_{\downarrow}(k''+ q) \Gamma_{\uparrow\downarrow}(k'',k';q)\,,
\end{multline}
where we introduced a momentum-frequency notation with $k=(\veck,i\omega_{n})$, $q=(\vecq,i\nu_{m})$ for the fermionic and bosonic variables, respectively, with $\omega_{n} = (2n + 1) \pi T$ and $\nu_{m} = 2m\pi T$ at temperature $T$. We denoted  $k$ and  $k'$  momentum and  frequency  of the incoming and outgoing particle carrying spin up and $q$ is the difference between the momentum-frequency variables of the particle and the hole. Vertex $\Lambda_{\uparrow\downarrow}^{eh}$ is irreducible with respect to simple electron-hole scatterings. That is, it cannot be disconnected by cutting two antiparallel lines. The reducible diagrams are summed in the Bethe-Salpeter equation~\eqref{eq:BS-eh}, see Fig.~\ref{fig:BSE-eh}. Analogously, the  Bethe-Salpeter equation with multiple scatterings  of two electrons with opposite spins, electron-electron channel, is in the same notation 
\begin{multline}\label{eq:BS-ee}
\Gamma_{\uparrow\downarrow}(k,k';q) = \Lambda_{\uparrow\downarrow}^{ee}(k,k';q) 
\\
 - \frac 1{\beta N}\sum_{k''} \Lambda_{\uparrow\downarrow}^{ee}(k,k'';q + k' - k'') G_{\uparrow}(k'') 
\\
\times G_{\downarrow}(q + k + k' - k'') \Gamma_{\uparrow\downarrow}(k'',k';q + k - k'') 
\end{multline}
 that is diagrammatically represented in Fig.~\ref{fig:BSE-ee}.

A singularity in the Bethe-Salpeter equation emerges  at the Fermi energy,  $\nu_{m}=0$, and in the spin-symmetric phase, $G_{\uparrow}= G_{\downarrow}$ with increasing the interaction strength at a vector $\vecq$. To simplify the reasoning we assume a homogeneous critical point with $\vecq = 0$. It is the electron-hole channel that is singular for the repulsive interaction and the electron-electron channel for the attractive coupling. We will investigate the repulsive case and a magnetic transition. That is, matrix 
\begin{equation}\label{eq:M-matrix}
M_{k,k'}= \beta N\delta_{k,k'} + \Lambda_{\uparrow\downarrow}^{eh}(k,k';0) G(k')^{2}
\end{equation}
has zero eigenvalue at a critical interaction strength. This critical point corresponds to the paramagnetic-ferromagnetic transition with a divergent  homogeneous magnetic susceptibility.  The antiferromagnetic transition would emerge at a vector $\mathbf{Q}$ for which $\epsilon(\veck + \mathbf{Q}) = -\epsilon(\veck)$. The magnetic susceptibility at zero magnetic field is defined 
\begin{multline}\label{eq:Susceptibility-local}
\chi = \frac{d m}{d h} = \frac 1N\sum_{\mathbf{i}}\left[\frac{d \langle \widehat{n}_{\mathbf{i}\uparrow}\rangle}{dh} - \frac{d \langle \widehat{n}_{\mathbf{i}\downarrow}\rangle} {dh}\right] = \frac 2N\sum_{\mathbf{i}}\frac{d \langle \widehat{n}_{\mathbf{i}\uparrow}\rangle} {dh} 
\\
= \frac 2{\beta N}\sum_{k}\frac{d}{dh} G_{\uparrow}(k) =- \frac 2{\beta N} \sum_{k} G(k)^{2}\left(1 - \frac{d\Sigma_{\uparrow}(k)}{dh} \right)  \ .
\end{multline}

The susceptibility diverges only if the derivative of the self-energy diverges and a spin-polarized self-energy emerges. This divergence must match the divergence in the Bethe-Salpeter equation~\eqref{eq:BS-eh} in the thermodynamically consistent approach. It is the case when the Ward identity, Eq.~\eqref{eq:WI-functional}, is fulfilled. It contains a functional-derivative from which it is mostly impossible to resolve the self-energy from a given two-particle vertex. We hence have to resort to approximations. To reach a \textit{qualitative agreement} between the symmetry breaking in the self-energy and the divergence in the vertex function we linearize the Ward identity with respect to the external magnetic field and introduce an approximate thermodynamic self-energy
\begin{equation}\label{eq:SigmaT-general}
\Sigma_{\sigma}^{T}(k) = \Sigma_{0} + \frac 1{\beta N}\sum_{k'}\Lambda^{eh}(k,k'; 0) G_{-\sigma}(k')\,,
\end{equation}
where we introduced an initial self-energy $\Sigma_{0}$ independent of the magnetic field and used a symmetrized irreducible vertex $\Lambda^{eh}(k,k'; 0)=(\Lambda_{\uparrow\downarrow}^{eh}(k,k'; 0) +\Lambda_{\downarrow\uparrow}^{eh}(k,k'; 0))/2 $.  This symmetrization does not change the Bethe-Salpeter equation at zero magnetic field. It only guarantees that the symmetric vertex is an even function of the magnetic field, since generally $\Lambda_{\downarrow\uparrow}^{eh}(k,k'; 0) = \Lambda_{\uparrow\downarrow}^{eh}(-k,-k'; 0)$. Equation~\eqref{eq:SigmaT-general} is a linearized form of the full functional identity in that only a linear response to the external magnetic field has been taken into account in solving Eq.~\eqref{eq:WI-functional}.   Equation~\eqref{eq:SigmaT-general} does not lead to the full dependence of the self-energy on the external magnetic field, since vertex $\Lambda^{eh}_{\uparrow\downarrow}$ is only a partial (functional) derivative of the self-energy. To restore the full dependence one has to add the other irreducible vertices. The contribution from vertex $\Lambda^{ee}_{\uparrow\downarrow}$ compensates the contribution from vertices $\Lambda^{ee}_{\uparrow\uparrow}$ and $\Lambda^{eh}_{\uparrow\uparrow}$  at the critical point of the magnetic transition and their dependence on the magnetic field can be neglected within linear response. Our objective is to reach a qualitative agreement between the one and two-particle functions in the critical region for which Eq.~\eqref{eq:SigmaT-general} is sufficient.  

Within the linear response we can neglect  dependence of vertex $\Lambda^{eh}$ on the magnetic field and can represent the derivative of the self-energy from Eq.~\eqref{eq:SigmaT-general} as 
\begin{multline}\label{eq:SigmaT-derivative}
\frac{d \Sigma^{T}_{\uparrow}(k)}{d h} = \frac 1{\beta N} \sum_{k'} \frac {d}{dh}\left[\Lambda^{eh}(k,k';0)G_{\downarrow}(k')\right] 
\\
= - \frac 1{\beta N} \sum_{k'} \Lambda^{eh}(k,k';0)\frac {d}{dh} G_{\uparrow}(k')
\\
= \frac 1{\beta N} \sum_{k'} \Lambda^{eh}(k,k';0) G(k')^{2}\left[ 1 - \frac{d\Sigma^{T}_{\uparrow}(k')}{dh}\right] \,,
\end{multline}
where we used the symmetry $d G_{\downarrow}/d h = - d G_{\uparrow}/d h$ at $h=0$.

The derivative of the self-energy is then determined generally from a matrix (integral) equation
\begin{subequations}\label{eq:DSigma}
\begin{multline}\label{eq:dSigma-integral}
\frac 1{\beta N} \sum_{k'}\left[ \beta N\delta_{k,k'} + \Lambda^{eh}(k,k';0) G(k')^{2}\right]\frac{d\Sigma^{T}_{\uparrow}(k')}{dh} 
\\
=  \frac 1{\beta N} \sum_{k'} \Lambda^{eh}(k,k';0) G(k')^{2}
\end{multline}
that has the same integral kernel as the Bethe-Salpeter equation~\eqref{eq:BS-eh}. We can hence represent its solution via the full vertex as  
\begin{equation}\label{eq:dSigma-Gamma}
\frac{d\Sigma^{T}_{\uparrow}(k)}{dh} = \frac 1{\beta N}\sum_{k'}\Gamma(k,k';0)G(k')^{2} \ .
\end{equation}
\end{subequations}
The derivative of the thermodynamic self-energy has the same divergence as the Bethe-Salpeter equation~\eqref{eq:BS-eh} determined by zero eigenvalue of matrix $M_{n,n'}$ from Eq.~\eqref{eq:M-matrix}. It means that the singularity in the Bethe-Salpeter equation is accompanied by a symmetry breaking in the self-energy if the one-electron propagators are renormalized with the thermodynamic self-energy from Eq.~\eqref{eq:SigmaT-general}. In this way qualitative thermodynamic consistency is achieved and quantum criticality of the two-particle function coincides with the critical behavior of the derivative of the self-energy with respect to the symmetry-breaking field.

Having introduced a thermodynamic self-energy to be used in the two-particle functions, we have to elucidate the role of the self-energy from the Schwinger-Dyson equation that is an exact dynamical equation resulting from the functional many-body Schr\"odinger equation.\cite{Mahan:2000aa} Its explicit form is
\begin{multline}\label{eq:SDE-general}
\Sigma_{\sigma}(k) = \frac U{\beta N}\sum_{k'}G_{-\sigma}(k')
\left[ e^{i\omega_{n'}0^{+}} - \frac 1{\beta N}\sum_{k''}  G_{\sigma}(k'') 
\right. \\ \left.\phantom{\frac 12}
 \times G_{-\sigma}(k''+ k'- k )   \Gamma_{\sigma-\sigma}(k'',k;k - k') \right]\,.
\end{multline}
This self-energy differs from the thermodynamic one $\Sigma^{T}_{\sigma}$ from Eq.~\eqref{eq:SigmaT-general} in approximate treatments when the full two-particle vertex $\Gamma_{\sigma-\sigma}$ is determined from the Bethe-Salpeter equation~\eqref{eq:BS-eh} with the generating irreducible vertex $\Lambda^{eh}_{\sigma-\sigma}$. The existence of two self-energies is a consequence of the uniqueness of the two-particle vertex. The concept of two-self-energies is not unusual and is used for instance in the local-moment approach of Logan and his group.\cite{Logan:1998aa,Logan:2015aa} One of the self-energies must then be used only as an auxiliary function. In our case it is the thermodynamic one, $\Sigma^{T}_{\sigma}$, that is used in the one-particle self-consistency to renormalize the one-particle propagators in the perturbation theory. The physical self-energy is then $\Sigma_{\sigma}$ from Eq.~\eqref{eq:SDE-general} where the one-electron propagators use the thermodynamic self-energy, $G_{\sigma}(k) = [i\omega_{n} + \mu + \sigma h - \epsilon(\veck) - \Sigma^{T}_{\sigma}(i\omega_{n},\veck)]^{-1}$. We call the physical self-energy from Eq.~\eqref{eq:SDE-general} spectral, since it determines the spectral and dynamical properties of the equilibrium state. It has a richer dynamical structure than the thermodynamic one and, what is most important, it generates qualitatively the same thermodynamic behavior with he same critical point in the susceptibility.  We demonstrate this on its derivative with respect to the magnetic field. The derivative of the spectral self-energy is  
\begin{widetext}
\begin{multline}\label{eq:SD-derivative}
\frac{d \Sigma_{\uparrow}(k)}{d h} = \frac U{\beta N} \sum_{k'}G(k')^{2}
 \left[ 1 - \frac{d\Sigma^{T}_{\uparrow}(k')}{dh}\right]\left\{1 - \frac 1{\beta N} \sum_{k''} G(k'')\left[G(k + k' - k'')
\phantom{\frac 12}
\Gamma(k'',k;k'- k'')+ G(k'+ k''-k)
\right.\right. \\ \left.\left.  \phantom{\frac 12}
\times\Big(\Gamma(k'',k;k'- k) - \Gamma(k',k;k''-k) \Big)\right] \right\}\,.
\end{multline}
Since the irreducible vertex $\Lambda^{eh}_{\uparrow\downarrow}$  depends on even powers of the magnetic field, the derivative of the full  vertex $\Gamma_{\uparrow\downarrow}$, defined in Eq.~\eqref{eq:BS-eh},  vanishes at $h=0$. 

The magnetic susceptibility at $h=0$ calculated from the Green function with the spectral self-energy then is
\begin{multline}\label{eq:chi-spectral}
\chi = - \frac 2{\beta N}\sum_{k}G(k - \Delta\Sigma(k))^{2}\left\{ 1 - \frac U{\beta N}\sum_{k'}G(k')^{2}\left[ 1 - \frac 1{\beta N}\sum_{k'''}\Gamma(k',k''';0)G(k''')^{2} \right] \left[1 - \frac 1{\beta N} \sum_{k''} G(k'')
\phantom{\frac 12}\right.\right. \\  \left.\left. \phantom{\frac 12}
\times\bigg(G(k + k'- k'')\Gamma(k'',k;k'- k'')+ G(k + k''- k)\left(\Gamma(k'',k;k'- k) - \Gamma(k',k;k''- k) \right)\bigg) \right]\right\}\,,
\end{multline}
\end{widetext} 
where $\Delta\Sigma(k) = \Sigma(k)- \Sigma^{T}(k)$.
It is clear that the magnetic susceptibility becomes critical  if the divergence in $\Gamma(k,k';q)$  at $q=0$ is independent of the incoming and outgoing fermionic energy-momenta $k$ and $k'$.  The physical self-energy then breaks its spin-reflection symmetry at the same critical point at which the full two-particle vertex from the Bethe-Salpeter equation in the singlet electron-hole channel, Eq.~\eqref{eq:BS-eh} has a pole.

\section{Approximate vertices: Two-particle self-consistency}
\label{sec:2P-SC}

\subsection{Hartree approximation}
\label{sec:2P-Hartree}

The fundamental object of our construction is the irreducible two-particle vertex generating singularity in the Bethe-Salpeter equation. It is the input into the theory and must be determined diagrammatically. Its simplest approximation is the bare interaction, that is $\Lambda^{eh}= U$. Then the thermodynamic self-energy is the Hartree one, $\Sigma^{T}_{\sigma} = Un^{T}_{-\sigma}$, while the spectral self-energy is determined from a ladder approximation in the singlet electron-hole channel
\begin{equation}\label{eq:Sigma-Hartree}
\Sigma_{\sigma}(k) = \frac U{\beta N}\sum_{q}\frac{G_{-\sigma}(k + q)}{1 + U\chi_{\sigma-\sigma}(q)}\,.
\end{equation} 
We denoted the electron-hole bubble $\chi_{\sigma-\sigma}(q) = (\beta N)^{-1}\sum_{k}G_{\sigma}(k + q) G_{-\sigma}(k)$. Both self-energies give a mean-field description of the magnetic critical behavior at weak coupling and high spatial dimensions.  If we want to develop approximations applicable also in strong coupling and low dimensions we must go beyond the simplest approximation and introduce a two-particle self-consistency. To improve upon the Hartree approximation by replacing the thermodynamic self-energy in the one-electron propagators by the spectral one from Eq.~\eqref{eq:Sigma-Hartree}, resulting in the so-called FLEX approximation,\cite{Bickers:1989aa} is a step in the wrong direction. It breaks the Ward identity and disconnects the symmetry breaking in the spectral self-energy from the critical point of the two-particle vertex of the Schwinger-Dyson equation. The correct procedure is to improve upon the irreducible vertex $\Lambda^{eh}$. To avoid spurious and unphysical singularities such as the Hartree one in low spatial dimensions we need to introduce a two-particle self-consistency. We use the parquet approach to do so.    

\subsection{Parquet equations}
\label{sec:PE-full}

The idea of the parquet approach is to use complementarity of reducible contributions from the respective Bethe-Salpeter equations, second terms on right-hand sides of Eqs.~\eqref{eq:BS-eh} and~\eqref{eq:BS-ee}. Or, equivalently, the sum of the irreducible vertices from which their common part, the vertex irreducible in both channels,  is subtracted, gives the full two-particle vertex. We then have the fundamental two-channel parquet equation      
\begin{equation}\label{eq:parquet-basic}
\Gamma_{\uparrow\downarrow}(k,k';q) = \Lambda_{\uparrow\downarrow}^{eh}(k,k';q)
 + \Lambda_{\uparrow\downarrow}^{ee}(k,k';q)  - U \,.
\end{equation}
We approximated the vertex irreducible in both channels by the bare interaction, which is called a parquet approximation. Excluding the full vertex $\Gamma_{\uparrow\downarrow}$ from Eqs.~\eqref{eq:BS-eh} and~\eqref{eq:BS-ee} by Eq.~\eqref{eq:parquet-basic} we obtain a set of parquet equations determining the irreducible vertices $\Lambda^{eh}_{\uparrow\downarrow}$ and $\Lambda^{ee}_{\uparrow\downarrow}$ for the given interaction strength $U$ and the one-particle propagators $G_{\sigma}$.  The interaction strength is an external parameter but the one-electron propagators must be related to the vertices from the parquet equations in a consistent theory.  The parquet approach aims at simultaneous renormalizations of the one- and two-particle functions. It was introduced into non-relativistic many-body problems by De Dominicis and Martin\cite{DeDominicis:1962aa,DeDominicis:1963aa, DeDominicis:1964aa,DeDominicis:1964ab} and later used by a number of authors for interacting Fermi\cite{Roulet:1969aa,Nozieres:1969aa,Weiner:1970aa,Weiner:1971aa,Bickers:1989aa,Bickers:1989ab,Bickers:1991aa,Bickers:1992aa,Janis:1999aa} as well as Bose systems.\cite{Jackson:1982aa,He:1993aa} Interest in the parquet construction of two-particle irreducible vertices has recently been renewed with the increasing numerical power allowing for numerical solutions of the full set the of parquet equations in specific situations.\cite{Yang:2009aa,Tam:2013aa,Valli:2015aa,Li:2016aa}    
 
We are not aiming at solving the parquet equations in their most complete form but rather to demonstrate on them  how to achieve qualitative thermodynamic consistency between the singularity in the two-particle vertex and the one-electron self-energy in proximity of  critical points in Bethe-Salpeter equations. We hence use the simplest form of the parquet theory with only two channels with scattering of singlet electron-electron (hole-hole) and electron-hole pairs. This minimal set of parquet equations contains the singularity of the full vertex and can hence be used to study quantum criticality in models with the Hubbard local interaction.

\subsection{Reduced parquet equations}
\label{sec:PE-reduced}

The problem of the unrestricted general parquet equations is that they cannot be formulated in real frequencies, since the analytic structure of the resulting two-particle vertices is unknown. The full set of parquet equations can be solved only numerically and hence the spectral properties of the solution of the vertex functions and the self-energy are not directly accessible.  Moreover, it is numerically hard  to go to very low temperatures and deep into the critical region in Matsubara formalism. There is yet a more important deficiency of the two-channel parquet equations with the bare interaction $U$ as its input.  They are unable to reach critical points in the Bethe-Salpeter equations and the strong-coupling Kondo critical regime in the single-impurity model. Divergence in the Bethe-Salpeter equation in one channel is transferred to the irreducible vertex in the other channel.  Convolutions of the divergent vertex induce new divergencies that must be compensated by corrections to the bare fully irreducible vertex. They are missing in the simplest version of the parquet equations.  When no compensation is present the two-particle self-consistency does not allow to reach the critical point.\cite{Janis:2006ab} It means that we either must go beyond the simplest parquet approximation with the bare interaction and introduce a dynamical fully irreducible vertex or we  simplify the parquet equations appropriately. 

An easier way is to slightly modify the structure of the two-particle self-consistency induced by the parquet equations.  We must proceed in such a way that the critical region of the underlying model is reachable. The singular channel in models with the repulsive interaction is the electron-hole one, Eq.~\eqref{eq:BS-eh} and the irreducible vertex $\Lambda^{ee}_{\uparrow\downarrow}$ is divergent at the critical point. We hence must avoid multiple convolutions of this singular function.
 We first replace the full vertex $\Gamma$ by the parquet equation, Eq.~\eqref{eq:parquet-basic}, on the left-hand side of the non-divergent Bethe-Salpeter equation~\eqref{eq:BS-ee}. Vertex $\Gamma$ on the right-hand side of this equation will be replaced by the non-singular irreducible vertex $\Lambda^{eh}$ so that not to enhance the singularity in the Bethe-Salpeter equation.  The irreducible vertex  $\Lambda^{ee}$ in this equation can be replaced in the leading order by its singular part, being the reducible vertex in the electron-hole channel $K^{eh}$. That is, the fully irreducible vertex (bare interaction) is subtracted from $\Lambda^{ee}$.  This replacement is necessary so that the proper balance between multiple scatterings from the electron-hole and electron-electron channels is achieved and the critical regime can be reached.  The critical behavior of the potentially singular Bethe-Salpeter equation has not been changed by this reduction. The suggested simplification should guarantee that when the fully irreducible vertex is replaced by the bare interaction in the parquet approach, the critical region can be reached.  
 
 The reduced parquet equations for the electron-hole irreducible and reducible vertices are after the introduced modifications
\begin{multline}\label{eq:BS-simplified-ee}
 \Lambda_{\uparrow\downarrow}^{eh}(k,k';q) = U 
 \\
 -  \frac 1{\beta N}\sum_{k''} K_{\uparrow\downarrow}^{eh}(k,k'';q + k' - k'') G_{\uparrow}(k'') 
 \\
 \times G_{\downarrow}( q + k + k'- k'')
  \Lambda_{\uparrow\downarrow}^{eh}(k'',k';q + k - k'') \ 
  \end{multline}
  and
\begin{multline}\label{eq:BS-simplified-eh}
K^{eh}_{\uparrow\downarrow}(k,k';q) =   - \frac 1{\beta N}\sum_{k''} \Lambda_{\uparrow\downarrow}^{eh}(k,k'';q)G_{\uparrow}(k'')
\\
\times  G_{\downarrow}(k''+ q) \left[K^{eh}_{\uparrow\downarrow}(k'',k';q) 
+ \Lambda^{eh}_{\uparrow\downarrow}(k'',k';q)\right] \ .
\end{multline}
These equations still contain the necessary two-particle self-consistency needed for a qualitatively correct description of the critical behavior of the full two-particle vertex $\Gamma_{\uparrow\downarrow}^{eh}$. It is important that this simplification does not change the structure of the poles in the Bethe-Salpeter equations.  Equations~\eqref{eq:BS-simplified-ee} and~\eqref{eq:BS-simplified-eh} are a generalization,  formalization of the simplified parquet equations introduced earlier in Refs.~\onlinecite{Janis:2007aa,Janis:2008aa} and used to describe the Kondo regime in SIAM.  

One can observe an important simplification in the reduced parquet equations. If we relabel the bosonic variables $q \to  q + k + k'$ in Eq.~\eqref{eq:BS-simplified-ee} and use it as the bosonic variable in vertex $ \Lambda_{\uparrow\downarrow}^{eh}$ we find that this  vertex does not explicitly depend on the outgoing fermionic variable $k'$.  We obtain a new representation of the reduced  parquet equations 

\begin{multline}\label{eq:Lambda-reduced}
 \Lambda(k;q) = U 
 - \frac 1{\beta N}\sum_{k''} K(k,k''; q - k - k'' ) 
 \\
 \times G(k'') G( q - k'')  \Lambda(k'';q)
 \end{multline} 
 and
\begin{multline}  \label{eq:K-reduced}
 K(k,k';q) 
 = - \frac 1{\beta N}\sum_{k''} \Lambda(k; q + k + k'')  G(k'')
 \\
 \times  G(q + k'')\left[ K(k'',k'q) 
 + \Lambda(k'';q + k''+ k') \right]\,,
 \end{multline}
where we skipped the upper index $eh$ at the vertex functions.  
 
With the two different bosonic variables used in the parquet equations we have two ways to represent the full two-particle vertex.  We either have
\begin{subequations}
\begin{align}\label{eq:Gamma-eh}
\Gamma(k,k';q) &= \Lambda(k;q) 
+ K(k,k';q - k - k') \ ,
\end{align}
or 
\begin{align}\label{eq:Gamma-ee}
\Gamma(k,k';q)  &= \Lambda(k;q + k + k')
 + K(k,k';q)  \ ,
\end{align}\end{subequations}
depending on whether it is more convenient to use the conserved momentum-frequency in the electron-electron channel, Eq.~\eqref{eq:Gamma-eh} or in the electron-hole channel, Eq.~\eqref{eq:Gamma-ee}.

The thermodynamic self-energy for the reduced parquet equations in the new notation is
\begin{equation}
\Sigma^{T}(k) = \Sigma_{0} + \frac 1{\beta N}\sum_{k'}\Lambda(k;k + k')  G(k')\ .
\end{equation}
The matrix equation for the derivative of the thermodynamic self-energy is  identical with that of the full parquet equations, see Eqs.~\eqref{eq:DSigma}, 
\begin{multline}\label{eq:dSigma-integral2}
\frac 1{\beta N} \sum_{k'}\left[ \beta N\delta_{k,k'} + \Lambda(k,k + k') G(k')^{2}\right]\frac{d\Sigma^{T}_{\uparrow}(k')}{dh} 
\\
=  \frac 1{\beta N} \sum_{k'} \Lambda(k;k + k') G(k')^{2} \ .
\end{multline}
Using the solution of Eqs.~\eqref{eq:K-reduced} and ~\eqref{eq:Lambda-reduced} we can represent the derivative via the vertex functions
\begin{multline}
\frac{d\Sigma^{T}_{\uparrow}(k)}{dh} 
\\
=   \frac 1{\beta N} \sum_{k'} \left[\Lambda(k,;k + k') 
+ K(k,k';0)\right]  G(k')^{2} \ .
\end{multline}
We further use Eq.~\eqref{eq:dSigma-Gamma} to represent the magnetic susceptibility calculated from the thermodynamic self-energy
\begin{multline}
\chi^{T} = - \frac 2{\beta N} \sum_{k}G(k)^{2}\left\{1 - \frac 1{\beta N}\sum_{k'}\left[ \Lambda(k;k + k')  \phantom{\frac 12}
\right.\right. \\ \left.\left. \phantom{\frac 12}
 + K(k,k';0)\right] G(k')^{2}\right\} \ .
\end{multline} 

Finally, the physical (spectral) self-energy has the following representation
\begin{multline}\label{eq:SD-Matsubara-new}
\Sigma_{\sigma}(k) =  \frac U{\beta N}\sum_{k'}G_{-\sigma}(k') \left[ e^{i\omega_{n}0^{+}} 
- \frac 1{\beta N}\sum_{k''}  G_{\sigma}(k'') 
\right. \\ \left.
 \times G_{-\sigma}(k' + k''- k) 
\left(\Lambda(k'';k'+ k'') \phantom{\frac 12} 
\right.\right. \\ \left.\left. \phantom{\frac 12}
+ K(k'',k;k'- k'') \right)   \right] \ .
\end{multline}
The magnetic susceptibility at zero magnetic field calculated from the Green function with the spectral self-energy was given in Eq.~\eqref{eq:chi-spectral}.

\section{Single-impurity Anderson model at half filling}
\label{sec:SIAM}

The simplest example of quantum criticality of correlated electrons is the single-impurity Anderson model. There is no critical point in this model for finite interaction strengths but the strong-coupling Kondo regime is a critical region of a metal-insulator transition at infinite interaction. A consistent description of this limit demands that the Kondo scale (temperature) determined from the one-particle spectral function, the width of the quasiparticle peak, or the two-particle magnetic susceptibility is qualitatively the same that saturates at the Kondo temperature. To reach this one needs a thermodynamically consistent description of quantum criticality. 

The Anderson model is described by the Hamiltonian 
\begin{multline}\label{eq:H-SIAM}
  \widehat{H}_{SIAM} = \sum_{{\bf k}\sigma} \epsilon({\bf k}) c^{\dagger}_{{\bf k}\sigma}   c^{\phantom{\dagger}}_{{\bf
k}\sigma}   +  \sum_{{\bf
k}\sigma}\left(V^{\phantom{*}}_{{\bf   k}}c^{\dagger}_{{\bf k}\sigma} d^{\phantom{\dagger}}_\sigma  + H.c.\right)
\\
+\  E_d\sum_\sigma  d^{\dagger}_{\sigma}d_{\sigma}   + U\widehat{n}^d_\uparrow\widehat{n}^d_\downarrow \,, 
\end{multline}
where $d^{\phantom{\dagger}}_\sigma$ and $d^{\dagger}_\sigma$ are annihilation and creation operators of the impurity electrons and $\widehat{n}^d_\sigma = d^{\dagger}_\sigma d^{\phantom{\dagger}}_\sigma$. We integrate over the itinerant degrees of freedom and introduce an energy scale on the impurity  
$\Delta(\epsilon) = \pi \sum_{\bf k} |V_{\bf k}|^2 \delta(\epsilon -
\epsilon({\bf k})) \doteq \Delta
$. Then only local degrees of freedom become relevant with an effective grand partition function represented via a functional integral over Grassmann variables
\begin{multline}
\mathcal{Z} = \int \mathcal{D}\psi \mathcal{D}\psi^* \exp\left\{\sum_n
\psi^*_n  \left[G_{0}(i\omega_{n})\right]^{-1} \psi_n 
\right. \\ \left.
 -\ U\int_0^\beta d\tau\ \widehat{n}^d_\uparrow(\tau)
\widehat{n}^d_\downarrow(\tau) \right\}
\end{multline}
 with $\left[G_{0}(i\omega_{n})\right]^{-1}= i\omega_{n} + \mu - E_{d} + i\Delta \mathrm{sign} \omega_{n}$.
 
We use our general construction of thermodynamically consistent approximations to obtain qualitatively the same Kondo scale from the spectral function and the local magnetic susceptibility.  The $\Phi$-derivable (FLEX) approximations do not lead to the Kondo scale at all.\cite{Hamann:1969aa} Apart from numerical methods based on the numerical renormalization group\cite{Krishna-murthy:1980aa} and the functional renormalization group,\cite{Karrasch:2008aa,Streib:2013aa} there are no analytic approaches that would predict the correct linear dependence of the exponent of the Kondo scale on the interaction strength.  We show that our general construction applied on SIAM does the job.         

\subsection{Effective-interaction approximation}
\label{sec:EIA}

We need to apply a two-particle self-consistency from the parquet approach to describe qualitatively correctly the strong-coupling limit of SIAM.  The reduced parquet equations do not generally allow for analytic continuation to real frequencies and must be solved only numerically in the Matsubara formalism.  To allow for an analytic representation of the vertex and spectral functions we simplify the reduced parquet equations in that we take into account only the potentially divergent fluctuations and neglect the bounded ones and replace them with constants.\cite{Janis:2007aa,Janis:2008aa} 

The fermionic variables are less important in the two-particle vertices, since their fluctuations are non-critical and are summed over in physical quantities. The irreducible vertex $\Lambda$ does not depend  on the outgoing frequency, since the fully irreducible vertex is the static bare interaction. We further neglect  its weak dependence  on the incoming fermionic frequency.  The same we do for vertex $K$.  This approximation should work well at low temperatures where we expect quantum critical behavior to emerge.  Since vertex $\Lambda$ is non-singular we can moreover neglect its dependence on the bosonic frequency. That is, we neglect all finite (non-critical) fluctuations and keep dynamical only the critical ones. We hence make the following replacements     
\begin{subequations}
\begin{align}
\Lambda_{\sigma}  &  = \Lambda(0;  0) \,, \\
K( i\nu_{m})  &  = K(0,0;i\nu_{m})\,.
\end{align}   \end{subequations}
This reduction allows us to find the vertex functions explicitly in the whole plane of complex frequencies. A similar reduction was already used in SIAM and led to a qualitatively correct Kondo scale in the spectral self-energy.\cite{Janis:2007aa,Janis:2008aa} What is done differently here is the renormalization of the one-particle propagators used in the equations defining the vertex functions. 

The reduced vertices at low temperatures in this approximation are determined from algebraic equations in form of linear fractions
\begin{align}\label{eq:Lambda-zero}
\Lambda & = \frac{U}{1 + \psi} \ ,
\\ \label{eq:K-zero}
K(E) & = -\frac{\Lambda^{2}\phi(E)}{1 + \Lambda\phi(E)}
\end{align}
with
\begin{multline}
\phi(E)
 = -  \int_{-\infty}^{\infty} \frac{d\omega}{\pi } f(\omega) \left[G(\omega ) 
 \right. \\ \left. 
 +\  G(\omega - E) \right] \Im G(\omega ) \ 
\end{multline}  
and
 \begin{equation}
 \psi =   \int_{-\infty}^{\infty} \frac{d\omega}{\pi }b(\omega)  \Im\left[K^{*}( - \omega)G(\omega) G^{*}(-\omega)\right] \,, 
 \end{equation}  
 where asterisk denotes complex conjugation and $f(\omega)$ and $b(\omega)$ are Fermi and Bose distribution functions.. The thermodynamic self-energy becomes a single real constant 
\begin{equation}
\Sigma^{T}(E) = \Sigma_{0} - \Lambda \int_{-\infty}^{\infty} \frac{d\omega}{\pi}f(\omega)\Im  G(\omega)  = \Sigma_{0} + \Lambda n^{T} \,,
\end{equation}  
where $n^{T}$ is the particle density calculated from the Green function with the thermodynamic self-energy and $\Sigma_{0} = (U - \Lambda)/2$.

The spectral self-energy is
\begin{multline}\label{eq:Sigma-sp}
\Sigma(E) = - U \int_{-\infty}^{\infty}\frac{d\omega}{\pi} \left\{ \frac{f(\omega)\Im G(\omega)}{1 + \Lambda\phi(\omega - E)}   
\right. \\ \left.
- b(\omega)  G(E + \omega) \Im\left[ \frac{1}{1 + \Lambda\phi(\omega)}\right]
\right\} \,.
\end{multline}

To find a representation of the derivative of the spectral self-energy and the magnetic susceptibility we introduce an auxiliary function
\begin{multline}\label{eq:Xn-def}
X(i\omega_{n}) = \frac 1\beta\sum_{m}\frac{G(i\omega_{n} + i\nu_{m})}{\left[1 + \Lambda\phi(i\nu_{m})\right]^{2}}\left[ G(i\omega_{n} + i\nu_{m})
\right. \\ \left.
\times\left(1 + \Lambda\phi(i\nu_{m})\right) + \Lambda\left(\kappa(i\nu_{m}) - \kappa(-i\nu_{m}) \right)\right]\,.
\end{multline}
with 
\begin{multline}
\kappa(E)= - \int_{-\infty}^{\infty}\frac{d\omega}{\pi}f(\omega) \left[G(\omega - E)^{2}
\right. \\ \left. 
+\ 2G(\omega + E)\Re G(\omega)\right]\Im G(\omega)\,.
\end{multline}
The derivative of the spectral self-energy then is 
\begin{equation}
\frac{d \Sigma_{\uparrow}(i\omega_{n})}{dh} = \frac{UX(i\omega_{n})}{1 + \Lambda\phi(0)}\,, 
\end{equation}
from which we obtain an explicit representation of the magnetic susceptibility at low temperatures
\begin{equation}\label{eq:chi-zero}
\chi = - \frac 2\beta \sum_{n}G(i\omega_{n} - \Delta\Sigma(i\omega_{n}))^{2}\left[ 1 - \frac{ UX(i\omega_{n})}{1 + \Lambda\phi(0)}\right] \, .
\end{equation}

It is easy to find the spectral representation of the derivative of the spectral self-energy via an auxiliary function
\begin{multline}\label{eq:XE-spectral}
X(E) = - \int_{-\infty}^{\infty}\frac{d\omega}{\pi} b(\omega)\left\{\frac{\Im\left[G(\omega)^{2} \right]}{1 + \Lambda\phi(\omega - E)} 
\right. \\ \left.
+ \frac{\Lambda\left(\kappa(\omega - E) - \kappa(E - \omega)\right)\Im G(\omega)}{\left[1 + \Lambda\phi(\omega - E)\right]^{2}}  
+\ \Lambda G(E + \omega)^{2} 
\right. \\ \left.
\times\Im\left[\frac{\phi(\omega)}{1 + \Lambda\phi(\omega)} \right] + \Lambda G(E + \omega)\Im\left[ \frac{\kappa(\omega) - \kappa^{*}(-\omega)}{\left[1 + \Lambda\phi(\omega)\right]^{2}}\right] \right\} \,.
\end{multline}

\subsection{Kondo critical behavior at zero temperature}
\label{sec:Kondo} 
 
 Kondo regime is reached when the denominator of vertex $K(\omega)$ approaches zero, with $a =1 + \Lambda\phi(0)\ll 1$ defining the Kondo dimensionless scale. In this strong-coupling limit we can use a low-frequency polar decomposition
 \begin{equation}\label{eq:K-pole}
 K(\omega) \doteq \frac{\Lambda}{1 + \Lambda \phi(0) -  i\Lambda\phi'\omega }\,,
 \end{equation}
 where the frequency variable is tacitly assumed to be taken with an infinitesimal positive imaginary part when not said otherwise. Here 
 \begin{subequations}\begin{align}
 \phi(0) &= - \int_{-\infty}^{0}\frac{d\omega}{\pi} \Im\left[ G(\omega)^{2}\right] \ , \\
 \phi'& =\frac{ [\Im G(0)]^{2}}{\pi} \,.
 \end{align}\end{subequations}
We then obtain   
\begin{multline}
 \psi = - \Lambda\int_{-\infty}^{0}\frac{d\omega}{\pi} \Im\left[ \frac{ G(\omega)G^{*}(-\omega)}{a - i\Lambda\phi'\omega}\right]
 \\
\doteq \frac{ [\Im G(0)]^{2}\lvert \ln a \rvert}{\pi \phi' } = \lvert \ln a \rvert \ .
 \end{multline}  
The dimensionless Kondo scale then is
\begin{align}\label{eq:Kondo-exp}
a = e^{-U\rho_{0}} \,,
\end{align}  
where $\rho_{0}= -\Im G(0)/\pi = 1/\pi \Delta$ is the density of states at the Fermi energy. The prefactor in the exponent of the Kondo scale slightly differs from the exact one for the Lorentzian density of states. The Bethe-ansatz solution gives $\pi^{2}/8$. Be aware, however, that this prefactor is nonuniversal and depends on the form of the density of states. Our approximation reproduces the Kondo scale only qualitatively, that is, predicts linear dependence of the exponent on the interaction strength.   

This result was derived already earlier in Ref.~[\onlinecite{Janis:2007aa}], since the effective-interaction approximation is essentially equivalent to the simplified parquet equations used there. The equivalence holds, however, only at zero temperature at the charge and spin-symmetric situation where the self-energy is compensated  from symmetry reasons by the chemical potential. If we move away from half filling or break the spin-reflection symmetry, the two approaches differ. The former approach uses a static Hartree-like self-energy $\Sigma_{\sigma} = U n_{-\sigma}$ with the spin density $n_{\sigma}$ calculated with the full one-electron propagator and the spectral self-energy from the Schwinger-Dyson equation. Although there is no difference at the Kondo scale determined from the spectral function, there is a significant difference in the thermodynamic Kondo scale determined from the local magnetic susceptibility. The susceptibility with the thermodynamic self-energy, Eq.~\eqref{eq:Susceptibility-local}, is proportional to the inverse Kondo scale 
\begin{align}
\chi^{T} &\doteq  \frac 2a \int_{-\infty}^{0}\frac{d\omega}{\pi}\Im\left[ G(\omega)^{2}\right] ,
\end{align}
as well as the susceptibility using the spectral self-energy, Eq.~\eqref{eq:chi-zero}, with the auxiliary function $X(\omega)$, Eq.~\eqref{eq:XE-spectral},
\begin{align}
\chi &\doteq - \frac {2U}a \int_{-\infty}^{0}\frac{d\omega}{\pi}\Im\left[ G(\omega - \Delta\Sigma(\omega) )^{2}X(\omega)\right] \,.
\end{align}

 The magnetic susceptibility for the one-electron propagators using the Hartree self-energy reads
 \begin{equation}\label{eq:HF-susceptibility}
 \chi^{HF} =\frac{2 \displaystyle{\int_{-\infty}^{0}\frac{d\omega}{\pi}\Im\left[G(\omega - \Delta\Sigma(\omega))^{2} \left(1 - UX(\omega) \right) \right]}}{\displaystyle{ 1 + U^{2}\int_{-\infty}^{0}\frac{d\omega}{\pi}\Im\left[G(\omega - \Delta\Sigma(\omega))^{2} X(\omega) \right]}}
 \end{equation}
with no exponentially small Kondo scale. Although the Hartree self-energy in the one-electron propagators of the parquet equations reproduces qualitatively correctly the Kondo asymptotics of the spin-symmetric spectral function, the width of the Kondo resonance, it is unable to generate the Kondo scale in the magnetic susceptibility. Since the Hartree self-energy does not match the criticality in the two-particle vertex, there is no guarantee that the susceptibility from Eq.~\eqref{eq:HF-susceptibility} is free from unphysical instability. The denominator in the susceptibility is a decreasing function of the interaction strength and a spurious magnetic instability occurs  at a finite interaction strength
\begin{equation}\label{eq:HF-instability}
U^{2}\int_{\infty}^{0}\frac{d\omega}{\pi}\Im\left[G(\omega - \Delta\Sigma(\omega))^{2} X(\omega) \right] = - 1 \,.
\end{equation}    

The situation is not improved if  the self-energy is determined self-consistently from the Schwinger-Dyson equation. Such a construction fails  to reproduce the  exponentially small Kondo scale emerging either in the spectral or in the thermodynamic functions. A spurious magnetic instability, as in Eq.~\eqref{eq:HF-instability}, occurs as well at a finite interaction strength in FLEX-type approximations.

\section{Numerical results}
\label{sec:numeric}

Numerical solution of the full reduced parquet equations is unbiased but reachable only for rather high temperatures and not very strong interaction.  The effective-interaction approximation allow us to find an approximate analytic form of the Kondo asymptotics as  presented in Sec.~\ref{sec:Kondo}. We now present a full numerical solution of Eqs.~\eqref{eq:Lambda-zero} and~\eqref{eq:K-zero} at half filling and low temperatures. The thermodynamic self-energy is exactly compensated by the chemical potential  due to electron-hole symmetry,  $n^{T}= n= 1/2$ and $\mu - \Sigma_{0} = \Lambda/2$. The one-electron propagators are then the bare ones, $G(z) = [z + i\Delta \rm{sign}\Im z]^{-1}$. We set the energy scale $\Delta=1$. A solution to Eqs.~\eqref{eq:Lambda-zero} and~\eqref{eq:K-zero} can be reached even for rather strong interactions. Hence, we can trace forming of the exponential Kondo scale.     

\begin{figure}
\includegraphics[width=8.5cm]{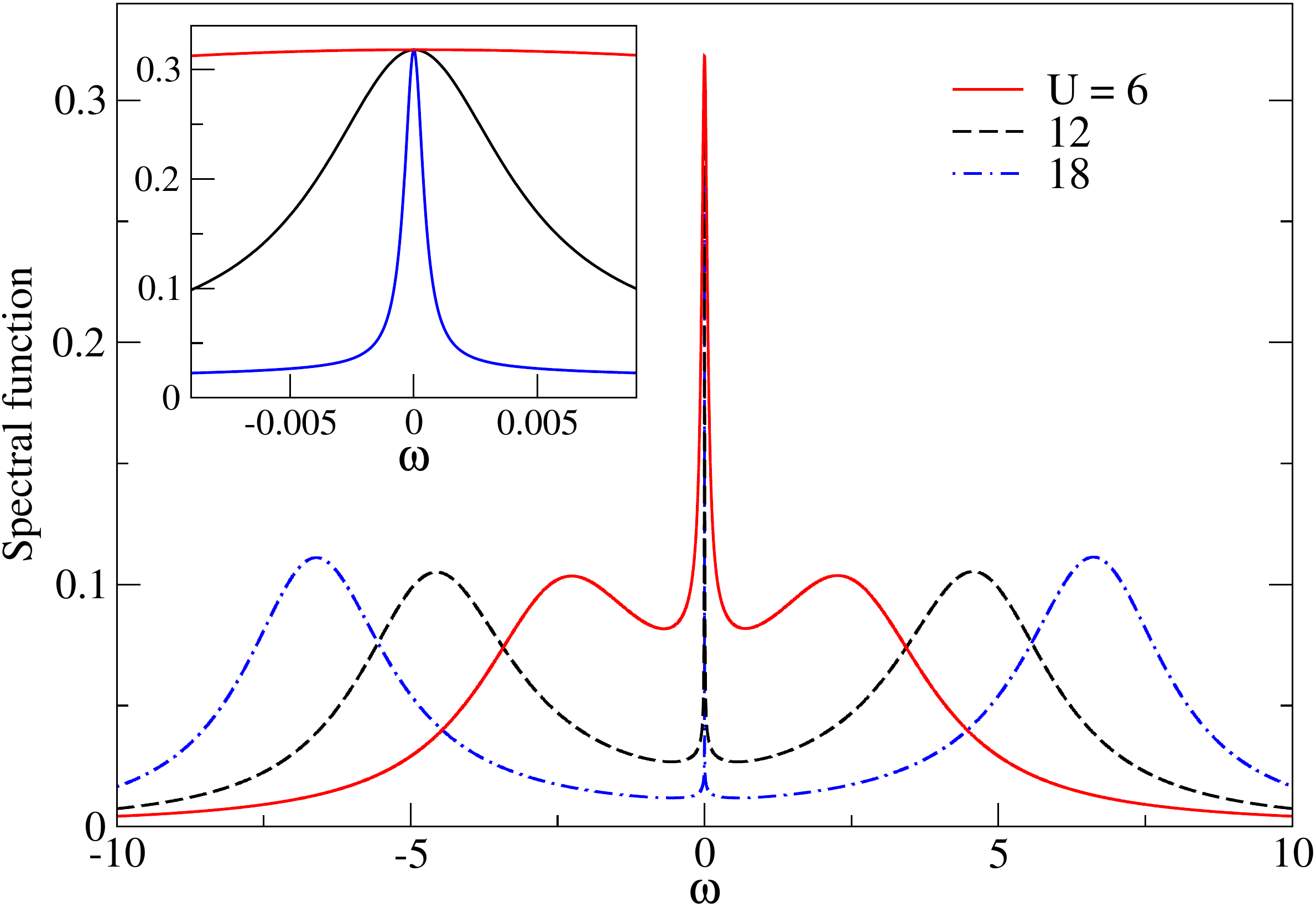}
\caption{(Color online)  Spectral function of SIAM at zero temperature and half filling calculated within the effective-interaction approximation. Formation and separation of the central quasiparticle and satellite peaks with increasing interaction strength is well reproduced. The inset is a  magnification of the central peak.\label{fig:spectral}}
\end{figure}

\begin{figure}
\includegraphics[width=8.5cm]{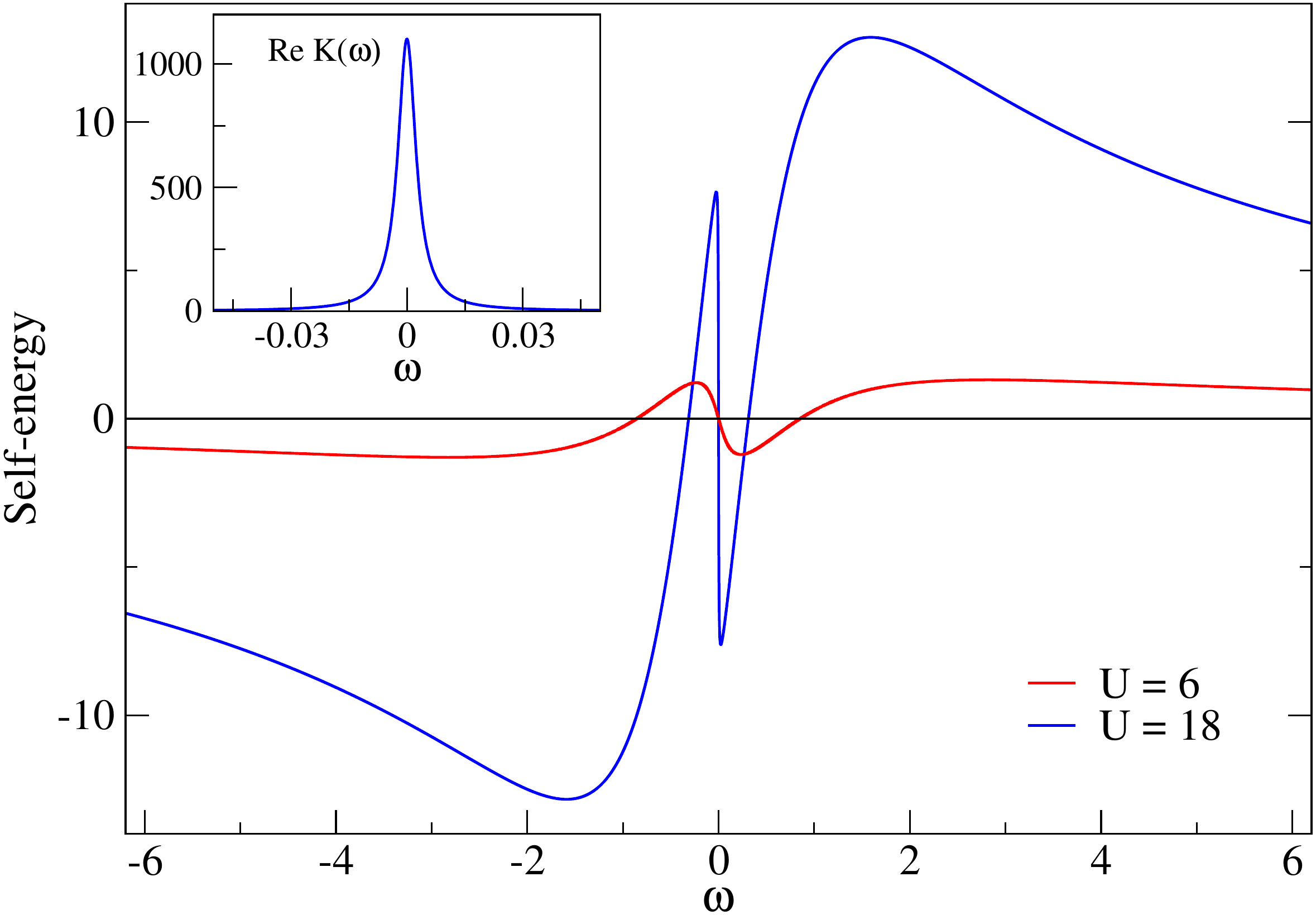}
\caption{(Color online) The real part of the spectral self-energy in the strong-coupling regime for the same setting as in Fig.~\ref{fig:spectral}. A broader peak beyond the Fermi-liquid regime near the Fermi energy is responsible for the high-energy satellite peak in the spectral function. The inset shows the real part of the singular two-particle vertex for $U=18$.   \label{fig:sigma}}
\end{figure}

\begin{figure}
\includegraphics[width=8.5cm]{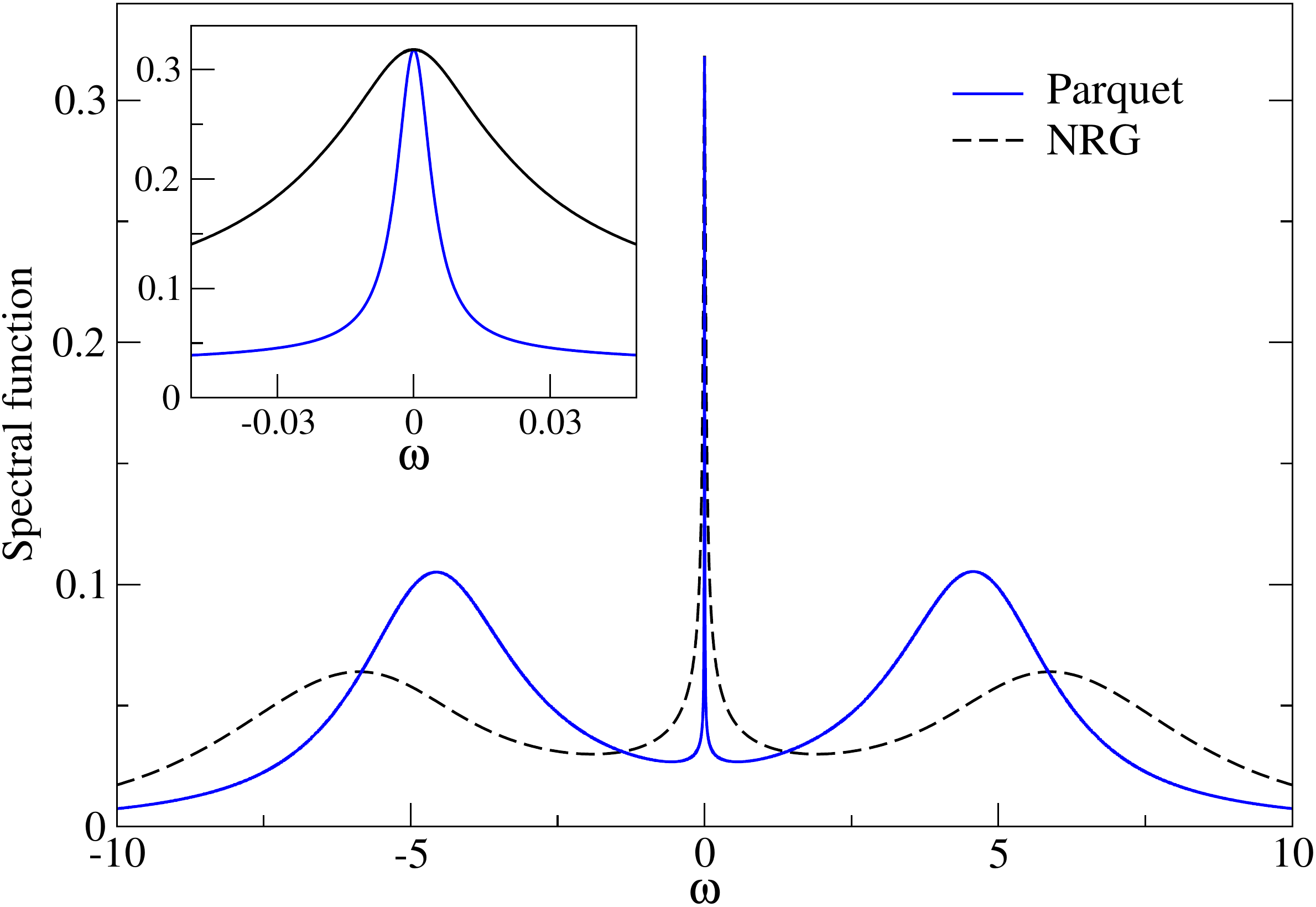}
\caption{(Color online)  Comparison of the zero-temperature spectral function of the impurity model  at half filling calculated  from the effective-interaction approximation to the parquet equations and NRG for $U= 8$. The inset is a magnification of the quasiparticle peak. \label{fig:spectral-nrg}}
\end{figure}

We plotted in Fig.~\ref{fig:spectral} the spectral function for several values of the interaction strength to demonstrate the formation of the low-frequency quasiparticle and high-frequency satellite peaks. The canonical three-peak structure of the spectral function can be deduced from the behavior of the real part of the self-energy, Fig.~\ref{fig:sigma}. The width of the central peak is determined by the slope, the derivative of the self-energy at the Fermi level. The two sharp peaks in the real-part of the self-energy delimit the Fermi-liquid domain. New peaks develop with increasing interaction beyond these limits. The height of this peak is decisive for formation of the satellite peaks in the spectral function. If $|\Re\Sigma(\omega)| > |\omega|$ then a new maximum in $\Im G(\omega)$ starts to develop around $\omega_{0} = \Re\Sigma(\omega_{0})$. The center of the satellite peaks is close to the atomic value at $\pm U/2$. It is slightly closer to the Fermi energy in the effective-interaction approximation. The height of the sharp peaks delimiting the region of the Fermi liquid determines how much the density of states is suppressed between the central and satellite peaks. It can vanish only for a finite bandwidth $w$ of the unperturbed energy spectrum if $|\Re\Sigma(\omega) - \omega| = w$ and simultaneously $\Im\Sigma(\omega)=0$. It cannot happen in SIAM with the Lorentzian density of states. To assess accuracy of our approximation we compared our spectral function with that from the Numerical Renormalization Group (NRG) in Fig.~\ref{fig:spectral-nrg}. The NRG data were obtained by the NRG Ljubljana code.\cite{NRGLjubljana}   We can see that the effective-interaction approximation gives a narrower quasiparticle peak and more pronounced satellite Hubbard bands.  

\begin{figure}
\includegraphics[width=8.5cm]{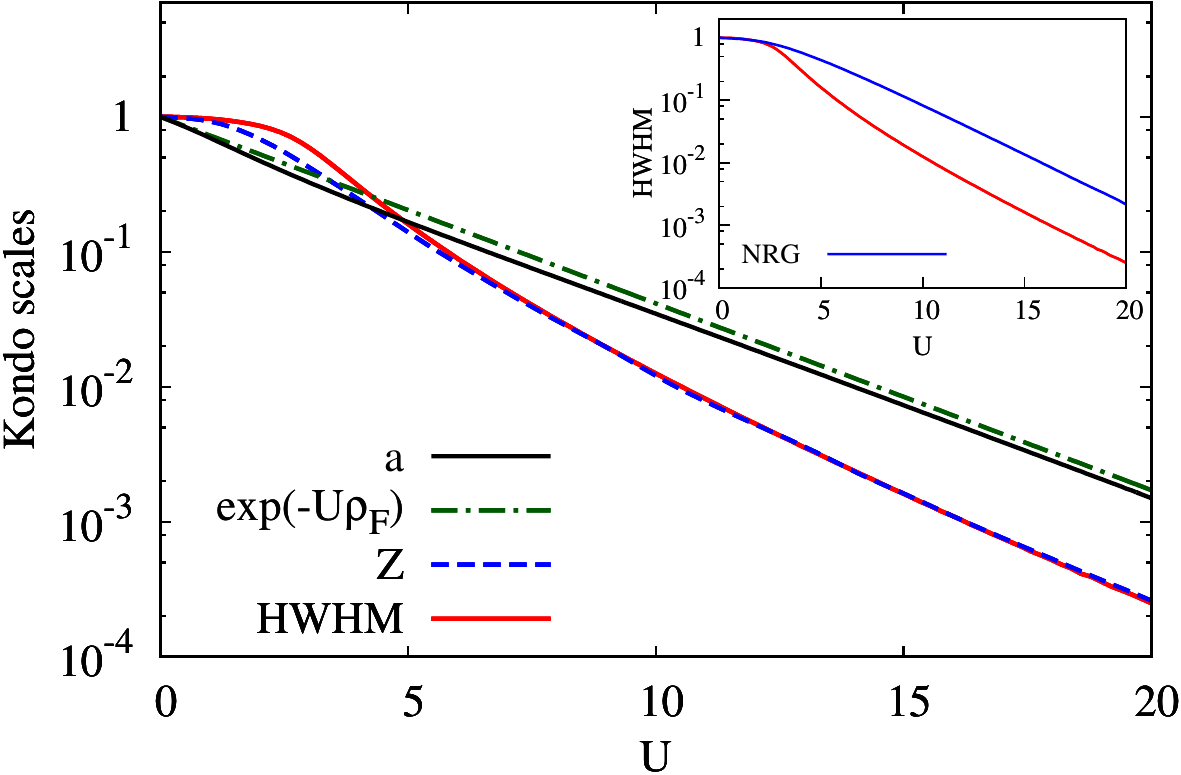}
\caption{(Color online) Various definitions of the Kondo scale, defined from the derivative of the self-energy ($Z$), half-width at half maximum of the central peak ($HWHM$),  from formula~\eqref{eq:Kondo-exp} ($\exp(-U\rho_{F})$), and the denominator of the singular two-particle vertex ($a$).  The inset shows the HWHM scale (red) compared with the NRG result (blue). \label{fig:Z}}
\end{figure}

The consistent description of quantum criticality should produce qualitatively the same critical behavior from different thermodynamically equivalent definitions. The measure of criticality in SIAM is the Kondo scale, or more precisely its asymptotic vanishing. The Green functions can be used to define the (dimensionless) Kondo scale in several ways. The primary definition is the critical asymptotics of the two-particle singular vertex $K(\omega)$ for $\omega=0$. We used the denominator of this vertex and defined the Kondo scale $a=1 + \Lambda \phi(0)$, see  Eq.~\eqref{eq:K-pole} or the asymptotics of vertex $\Lambda$ approaching its critical value, Eq.~\eqref{eq:Kondo-exp}. We can also extract the Kondo scale from the one-particle spectral function via the derivative of the self-energy $Z=1/(1 - \Sigma^{\prime}(0))$ or by taking the half-width of its quasiparticle peak at half maximum. These different definitions are compared in Fig.~\ref{fig:Z}. The first two scales coincide at strong coupling as well as do the latter two. The absolute values of the two scales differ numerically, since they contain different non-universal prefactors. They produce qualitatively the same exponential dependence of the Kondo scale on the bare interaction strength. Notice that the half-width of the quasiparticle peak is parallel to the exact Kondo scale in the inset of  Fig.~\ref{fig:Z}. It indicates that this definition of the Kondo scale fits quite well the exact result. Misplacement of our approximate curve is due to a difference in the pre-exponential factor caused by logarithmic corrections.   

\begin{figure}
\includegraphics[width=8.5cm]{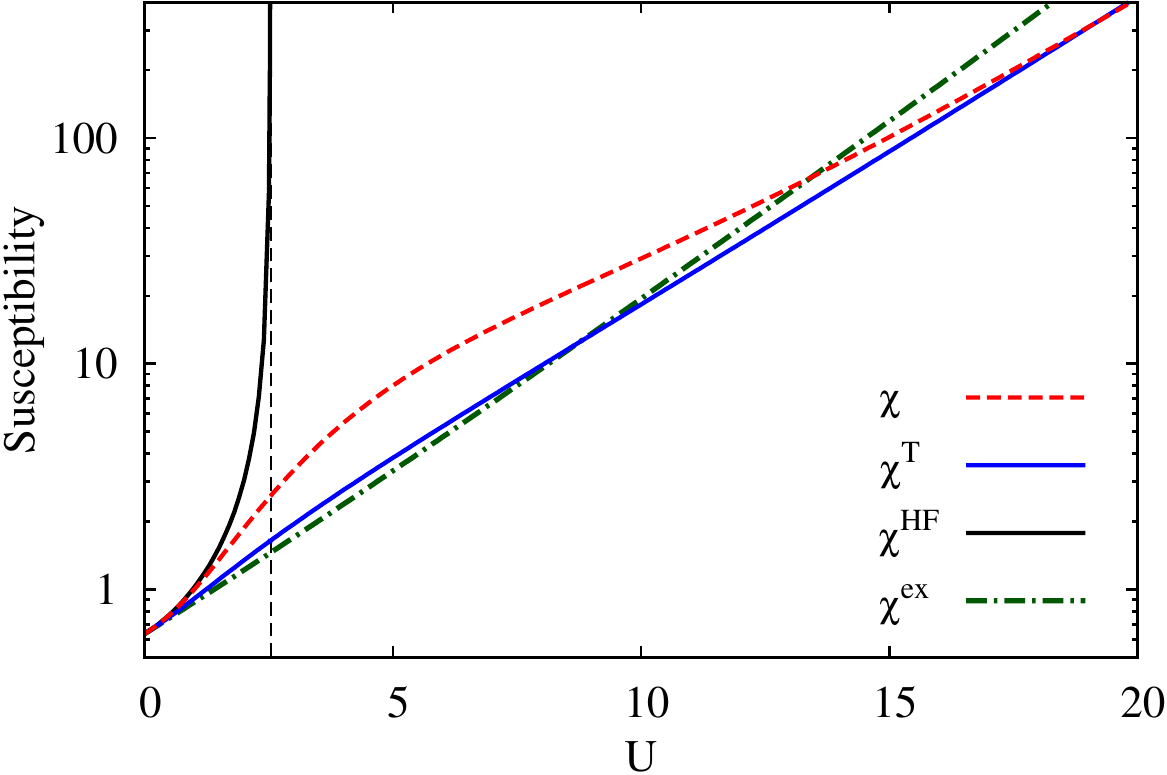}
\caption{(Color online) Zero-temperature magnetic susceptibilities calculated with the spectral self-energy ($\chi$), the thermodynamic self-energy ($\chi^{T}$), the Hartree-Fock one ($\chi^{HF}$), and the exact Bethe ansatz solution ($\chi^{ex}$). While the first two determine the same Kondo scale slightly differing from the exact result in strong-coupling, the third one displays a spurious magnetic instability. \label{fig:chi}}
\end{figure}

The major objective of the consistent theory of quantum criticality is equivalence of the critical behavior derived from the spectral and thermodynamic functions. In our case it is a qualitative equivalence of the Kondo scale defined either from the spectral function, critical behavior of vertex $K(\omega)$, or the local magnetic susceptibility.  Since we  have two self-energies in our approximation, there are also two susceptibilities. The auxiliary one derived from the spin-dependent propagators with the thermodynamic self-energy, $\chi^{T}$ copies the Kondo scale from the two-particle vertex.  The physical susceptibility $\chi$ is that derived from the propagators with the spectral self-energy.  We plotted the two susceptibilities in Fig.~\ref{fig:chi}. We can see that they coincide in the strong-coupling regime and lead to asymptotically the same Kondo scale that slightly differs in the slope from the exact Bethe ansatz result, due to a difference in the prefactor in the exponent of the Kondo scale.     

We plotted also the susceptibility derived from the propagators with the static Hartree self-energy in Fig.~\ref{fig:chi}. It is worth noting that this susceptibility differs from the Hartree approximation, since the particle density is calculated from the propagator with the full self-energy from the Schwinger-Dyson equation. Although this approximation with the Hartree self-energy correctly reproduces at half filling the spectral Kondo scale, it fails to do it in the magnetic susceptibility.\cite{Janis:2009ab} This approximation gets unstable and the local susceptibility diverges at a critical interaction $U_{HF}\approx 2.45$ before the Kondo regime can be reached. This example demonstrates how important it is to check stability of spin-symmetric solutions before its conclusions are accepted.     
  
\begin{figure}
\includegraphics[width=8.5cm]{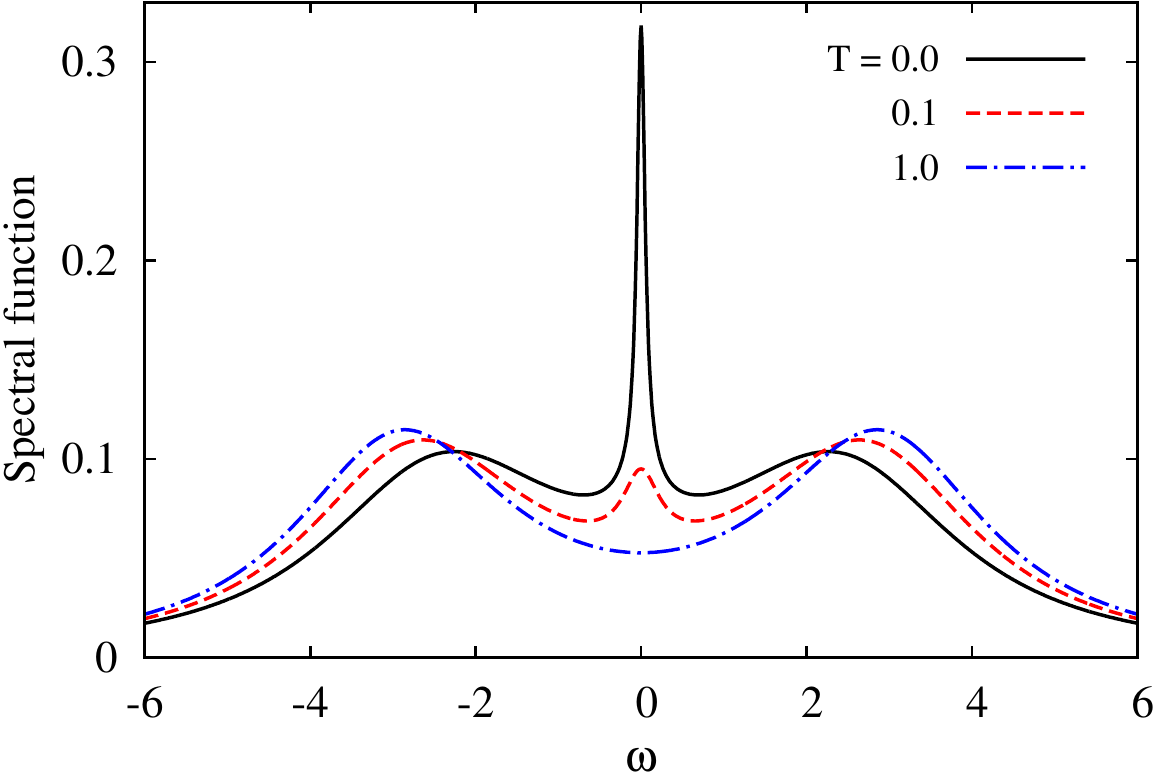}
\caption{(Color online) Formation of the quasiparticle peak in the spectral function at half filling with decreasing temperature for $U=6$.  \label{fig:dosT}}
\end{figure}

\begin{figure}
\includegraphics[width=8.5cm]{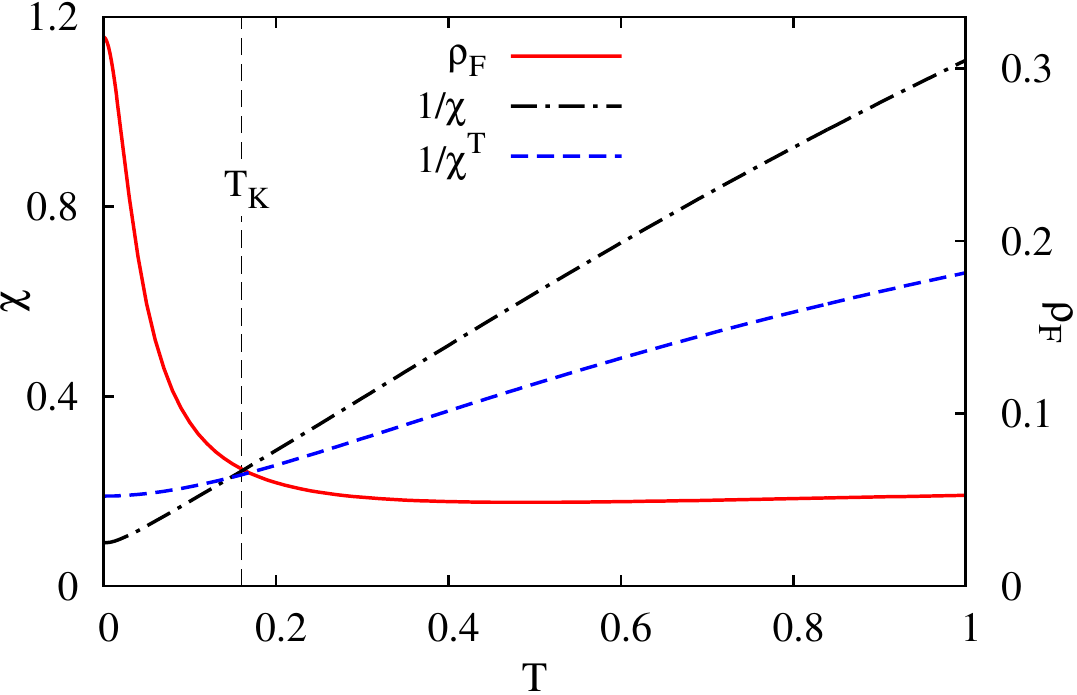}
\caption{(Color online) Temperature dependence of inverse susceptibilities $\chi^{T}$ and $\chi$  together with the density of states at the Fermi energy $\rho_{F}$ for $U=6$. Kondo temperature $ T_{K}=\sqrt{U/\pi \rho_{0}} \exp\{-\pi^{2}U\rho_{0}/8\}$ from the Bethe-ansatz solution was indicated for comparison.\label{fig:Tdep}}
\end{figure}

We continued the effective-interaction approximation to low non-zero temperatures to make an assessment of the temperature Kondo asymptotics. We plotted the spectral function for three temperatures for $U=6$ in Fig.~\ref{fig:dosT}. The three-peak structure becomes evident with the density of states at the Fermi energy approaching its zero-temperature, interaction-independent value, compare with Fig.~\ref{fig:Tdep}. The local magnetic susceptibilities $\chi$ and $\chi^{T}$ in Fig.~\ref{fig:Tdep} show the Curie  $T^{-1}$-behavior but saturate at zero temperature. The slopes and the limiting values differ from each other and also from the exact Kondo temperature $T_{K}$  due to different magnetic moment in different formulas.

\section{Discussion and conclusions}
\label{sec:discussion}

Thermodynamic consistency between one- and two-electron functions is guaranteed by Ward identities. It is, however, impossible to reconcile  the Ward identities with the Schwinger-Dyson equation in approximate treatments. Since we cannot guarantee that the single self-energy leads to  a single singular two-particle vertex we inverted the construction and set the two-particle irreducible vertex from the singular Bethe-Salpeter equation as a generating function. It means that we apply the two-particle diagrammatic expansion on a two-particle irreducible vertex that enters the Ward identity. The Ward identity is then used to determine a thermodynamic self-energy. The full functional Ward identity cannot generally be resolved and we resort to the Ward identity linearized in the symmetry-breaking field related to the critical point. Such a thermodynamic self-energy breaks its symmetry at the critical point of the two-particle function. This thermodynamic self-energy is then used in the renormalization of the one-electron propagators in the perturbation expansion of all physical quantities.

Approximate irreducible vertices cannot guarantee that the thermodynamic self-energy obeys the Schwinger-Dyson equation with the vertex satisfying the Bethe-Salpeter equation with the same irreducible vertex. We then have to distinguish two self-energies and give the thermodynamic self-energy an auxiliary role by renormalizing the one-electron propagators. The physical self-energy is then the one from the Schwinger-Dyson equation. The one-electron propagators in the Schwinger-Dyson equation are renormalized only with the thermodynamic self-energy and consequently the symmetry of the physical self-energy is broken at the critical point of the two-particle function.  In this way we achieved a qualitative consistency in the description of quantum criticality. 

We applied our construction on the single-impurity Anderson model where the critical behavior is the Kondo asymptotics with the critical point at infinite interaction. We used a parquet-type scheme with two scattering channels to demonstrate that the Kondo critical behavior can equivalently be determined from the spectral  as well as from thermodynamic functions. We showed that the Kondo exponential scale can be deduced either from the singular two-particle vertex, width of the quasiparticle peak in the spectral function, derivative of the spectral self-energy at the Fermi energy, and also from the local magnetic susceptibility. All the definitions lead to a qualitatively correct exponential Kondo scale. They reproduce its universal feature, that is, linear dependence of its logarithm on the bare interaction strength.            

To summarize, we presented a scheme how to generate approximations that produce a qualitatively consistent description of quantum criticality in correlated electron systems. On an example of the single-impurity model we demonstrated how to reduce the two-particle theory to analytically controllable approximations that allow for reaching the critical asymptotics of the relevant physical quantities. Qualitative consistency between different thermodynamically equivalent definitions is thereby guaranteed. This approximation can straightforwardly be generalized to lattice models and thus offers an affordable scheme of a consistent description of quantum criticality in models of correlated electrons. Consistency between the critical behavior in the one-electron spectral function and the magnetic susceptibility is of utter importance in dynamical mean-field theory and the description of the Mott-Hubbard metal-insulator transition.

\section*{Acknowledgments}
 Research on this problem was supported by Grant No. 15-14259S of the Czech Science Foundation.  We thank M. \v Zonda for providing us with the data from the NRG calculations. VJ thanks Ka-Ming Tam for discussions and comments on the numerical solution of the parquet equations and the Fulbright Commission for financing his stay at Louisiana State University, where part of the research on this problem was done.

\begin{appendix}
  \setcounter{equation}{0} \renewcommand{\thesection}{}
  \renewcommand{\theequation}{\Alph{section}.\arabic{equation}}

\section{Securing analyticity in the numerical calculation (zero temperature)}

The defining equations for the vertices at zero temperature can be reduced to contain only their imaginary parts with the Kramers-Kronig relations determining the corresponding real parts. We guarantee thereby analyticity of the functions for real frequencies also in the numerical evaluation when integrals are replaced by discrete sums.  The imaginary parts of the bubbles can be represented as 
\begin{multline}
\Im \phi(E) =  \int_{-|E|}^{|E|} \frac{d\omega}{\pi}\left[\theta(\omega)\theta(-E) - \theta(-\omega)\theta(E)\right] 
\\
\times \Im G(\omega) \Im G(\omega + E) \ ,
\end{multline} 
 \begin{multline}
 \Im \kappa(E) = - \int_{-|E|}^{|E|} \frac{d\omega}{\pi}\left[\theta(\omega)\theta(-E) - \theta(-\omega)\theta(E)\right] 
 \\
 \times \Im\left[G(\omega)^{2} \right]\Im G(\omega- E)\ .
   \end{multline}
All complex functions take their variables on the real axis as the limit from the upper complex half-plane.  

Analogously for the spectral self-energy
\begin{multline}
\Im\Sigma(E) =  U \int_{-|E|}^{|E|} \frac{d\omega}{\pi}\left[\theta(\omega)\theta(-E) - \theta(-\omega)\theta(E)\right] 
\\
\times \Im G(\omega + E) \Im\left[\frac{1}{1 + \Lambda\phi(\omega)}\right]\,.
\end{multline}
The corresponding real parts of the above analytic functions are obtained from the Kramers-Kronig relation
\begin{align}\label{eq:KK-real}
\Re X(E) & =  P\int_{-\infty}^{\infty} \frac{d\omega}{\pi} \frac{\Im X(\omega)}{\omega - E}\,,
\end{align}
where $P\int$ denotes the principal value of the integral.
\end{appendix}


\end{document}